\documentclass[%
reprint,
 amsmath,
 amssymb,
 aps,
prl,
 floatfix,
]{revtex4-1}
\usepackage{graphicx}
\usepackage{epsfig}
\usepackage{dcolumn}
\usepackage{bm}
\usepackage{hyperref}
\usepackage{amsmath}
\usepackage{longtable}
\usepackage{mathtools}
\usepackage{siunitx}
\usepackage{appendix}
\usepackage{todonotes}
\newcommand{\E}{\mathcal{E}}
\newcommand{\F}{\mathcal{F}}

\newcommand{\NB}{\mathrm{nb}}
\newcommand{\BB}{\mathrm{bb}}

\begin{document}

\title{Generalized threshold of longitudinal multi-bunch instability in
synchrotrons}

\author{Ivan Karpov\thanks{ivan.karpov@cern.ch} and Elena Shaposhnikova}
\affiliation{CERN, CH 1211 Geneva 23, Switzerland}

\date{\today}

\begin{abstract}

Beam stability is an essential requirement for particle accelerators.
Longitudinal coupled-bunch instabilities (CBI) are driven by beam interaction with long-range wakefields induced in the resonant structures with narrow-band impedance. Single-bunch loss of Landau damping (LLD) is mainly determined by short-range wakefields excited at any geometry change of the beam pipe (broadband impedance) and leads to undamped bunch oscillations. Up to now,  to define the threshold beam intensity or impedance, these two effects were evaluated separately.
We developed an approach to numerically solve the stability problem in a more general case and derived a new analytical threshold.
We have shown that LLD can modify the mechanism of multi-bunch instability and reduce the CBI threshold below the LLD threshold. This effect explains the existing observations in the CERN SPS and should be considered for future accelerators, such as HL-LHC, EIC, FCC, and others.

\end{abstract}
\maketitle
The interaction of charged particles with the accelerator environment (impedance) can result in the development of undamped or exponentially growing bunch oscillations. In the absence of synchrotron radiation damping, an accelerator design relies either on beam stabilization only by the natural frequency spread of individual particles, called Landau damping~\cite{Landau}, or on active damping systems in addition.

It is common in beam stability analysis to separate single- and multi-bunch collective effects since they are driven by short- and long-range wakefields, respectively~\cite{Lebedev1968, Sacherer2, Balbekov-Ivanov1986, Laclare1987, Chao, NG, HandBook2023}. 
For the longitudinal beam motion considered below, many impedance sources, related to the wake functions via the Fourier transform, can be modeled by the resonances
\begin{equation}
    \label{eq:bbr_nbr_impedance}
    Z (\omega)= \frac{R}{1+iQ\left(\omega/\omega_r-\omega_r/\omega\right)}
\end{equation}
with a shunt impedance $R$, quality factor $Q$, and resonant frequency $\omega_r = 2 \pi f_r$. The cases of $Q \sim 1$ and $Q \gg 1$ correspond to broadband (BB) and narrowband (NB) impedances, or short- and long-range wakefields, respectively.

\begin{figure*}[tbh!]
    \centering
    \includegraphics{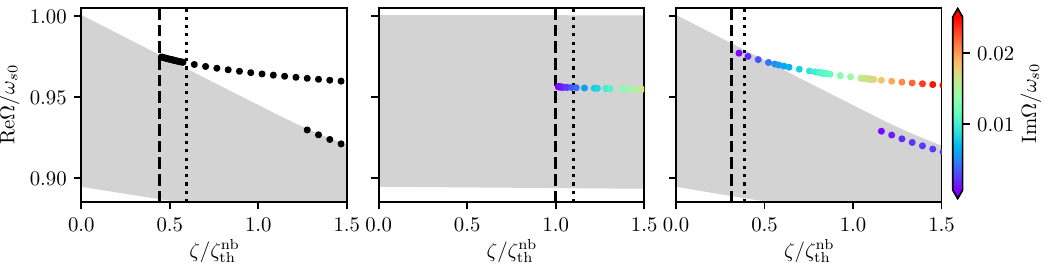}
    \caption{Evolution of coherent modes (black and colored circles) and incoherent frequency bands (gray) obtained with semianalytical code MELODY versus normalized intensity parameter $\zeta$ defined in Eq.~(\ref{eq:xi}). Left: LLD with BB impedance, no instability. Center: CBI with NB impedance. Right: CBI with BB and NB impedances. Thresholds found from exact Eq.~(\ref{eq:matrix_lebedev}) are shown with dashed lines while corresponding approximate solutions (\ref{eq:lld_threshold}), (\ref{eq:approximate_nb_threshold}), and (\ref{eq:general}) with dotted lines. Examples for nine bunches in the scaled LHC ring with parameters from Table~\ref{tab:LHC_parameters} and $k_\NB/h = 11/9$, $Q_\NB = 100$, $R_\NB = 37$~kOhm, $k_\BB/h = 5$, $Q_\BB = 1$, $R_\BB =30$~kOhm.}
    \label{fig: modes}
\end{figure*}

The synchrotron oscillations inside the bunch can be described as van Kampen modes~\cite{vKampen1, vKampen2, YHChin1983, burov2012van}. Their frequencies are modified by the presence of a BB impedance. At a certain intensity, the coherent mode moves outside the band of incoherent oscillation frequencies of individual particles leading to a loss of Landau damping (LLD), see Fig.~\ref{fig: modes}, left. Only the reactive (imaginary) part of the accelerator impedance is responsible for this effect, but the resistive impedance is needed to drive bunched-beam instabilities. 
A single-bunch instability resulting from the coupling of different coherent modes usually appears at bunch intensities significantly higher than the LLD threshold and is not discussed in this Letter. 
Coupled-bunch instability (CBI), leading to the coherent coupled motion of individual bunches, is possible if the decay time of the wakefield $2Q/\omega_r$ is longer than the bunch spacing. This instability takes place even if its coherent frequency lies inside the incoherent frequency band (Fig.~\ref{fig: modes}, center).

The threshold of CBI driven by a NB impedance was accurately calculated~\cite{Balbekov-Ivanov1986} using the matrix equation derived by Lebedev in 1968~\cite{Lebedev1968}. The LLD threshold in the longitudinal plane was first found in 1973 \cite{Sacherer2} and it was revised recently~\cite{IK2021}, using the Lebedev equation together with the approach~\cite{burov2012van} developed originally for the analysis of single-bunch instability~\cite{Oide_Yokoya1990}.
So far, the thresholds for these two effects were calculated separately, with few examples when the CBI growth rates were found in the presence of two impedance sources~\cite{Blaskiewicz2009, Burov2021}.

In this work, we evaluate the CBI threshold for two impedance types using the Lebedev equation. First, we consider them separately, and then we show that, if the LLD threshold is comparable to the CBI threshold, the instability in the presence of two impedances has a significantly reduced threshold and can even emerge below the LLD threshold. The mechanism of this multi-bunch instability is modified as well (Fig.~\ref{fig: modes}, right).  
We also derive an analytical expression for the general instability threshold and compare it with the results of self-consistent numerical calculations using code MELODY~\cite{MELODY} and macroparticle tracking simulations with code BLonD~\cite{BLOND2020,timko2022beam}.

Let us consider a beam of $M$ identical equidistant bunches, each containing $N_p$ particles with a charge $q$. 
We will use a coordinate system relative to the synchronous particle of the first bunch with the design energy $E_0$ and the rf phase $\phi_{s0}$, so that $\Delta E$ and $\phi$ are the particle energy and rf phase deviations. They are connected via equation, e.g.,~\cite{Chao},
\begin{equation}
    \frac{d\phi}{dt} \equiv \dot{\phi} = \frac{h^2\omega^2_0\eta}{\beta^2 E_0}\left(\frac{\Delta E}{h\omega_0} \right),  \label{eq:first_eq_motion}\nonumber
\end{equation}
where $\omega_0 = 2\pi f_0$, $f_0$ is the revolution frequency, $\beta$ the normalized particle velocity, $h$  the harmonic number, $\eta = 1/\gamma^2_\mathrm{tr} - 1/\gamma^2$ the slip factor, $\gamma$ the Lorentz factor with value $\gamma_\mathrm{tr}$  at transition energy. In this work, we analyze a single-rf case $V_\mathrm{rf}(\phi) = V_0\sin(\phi_{s0} + \phi)$, with $V_0$ being the rf voltage amplitude, but it can be extended to other rf waveforms.
Each particle performs synchrotron oscillations with energy $\mathcal{E}$ and phase $\psi$ 
\begin{align}
    \mathcal{E} &= {\dot{\phi}}^2/(2\omega^2_{s0}) +U_t(\phi), 
    \nonumber
    \\
    \psi &= \text{sgn}(\eta\Delta E)\frac{\omega_s(\mathcal{E})}{\sqrt{2}\omega_{s0}} \int_{\phi_\mathrm{max}}^\phi \frac{d\phi^\prime}{\sqrt{\mathcal{E} -U_t\left(\phi^\prime\right)}}, 
    \nonumber
\end{align}
where $\omega_{s0}$ is the angular oscillation frequency  of small-amplitude synchrotron oscillations in a bare single-rf potential $U_\mathrm{rf}$ and  $\omega_s(\mathcal{E})$ - oscillation frequency in total potential $U_t$, modified by intensity effects. 

Evaluation of beam stability is usually done in two steps.  First, a stationary solution should be found, e.g., by the iterative procedure~\cite{burov2012van} for some distribution function $\F= \F (\mathcal{E})$, 
bunch intensity, and impedance model. 
Below we consider the binomial family of the stationary distribution functions $\F(\E)\propto \left( 1 - \E/\E_{\max}\right)^{\mu}$, where $\E_{\max}$ and $\phi_{\max}$ are the maximum energy and phase of the synchrotron oscillations, and $\mu$ defines the bunch shape.
As a result, one obtains the line density $\lambda (\phi) = \int_{-\infty}^{\infty} \F(\E)d \dot{\phi}$ and  total potential 
\begin{equation}
    \label{eq:potential}
    U_t(\phi) = \frac{1}{V_0\cos{\phi_{s0}}}\int^\phi_{\Delta \phi_{s}} d\phi^\prime \, \left[V_t(\phi^\prime) - V_0\sin{\phi_{s0}} \right].
\end{equation}
The total voltage $V_t(\phi) = V_\mathrm{rf}(\phi) + V_\mathrm{ind}(\phi)$, contains contribution from the stationary beam-induced voltage
\[V_\mathrm{ind}(\phi)= - q M N_p \, h \, \omega_0 \sum_{k=-\infty}^{\infty} Z(k\omega_0) \lambda_k  e^{i \frac{k}{h} \phi},\]
with $k=f/f_0$ and $\lambda_k$ being the Fourier harmonics of the  line density 
normalized by $MN_p$
\begin{equation}
    \label{eq:line_density_harm}
    \lambda_k = \frac{1}{2\pi h} \int_{-\pi h}^{\pi h}d\phi \; \lambda(\phi) e^{-i \frac{k}{h} \phi},
\end{equation}
which are non-zero only for $k=pM, \; p=0,\pm 1, ...$
The synchronous phase shift due to intensity effects $\Delta \phi_s$ satisfies the relation $V_0 \sin{\phi_{s0}} = V_0\sin(\phi_{s0} + \Delta \phi_s) + V_\mathrm{ind}(\Delta \phi_s)$.

To proceed with the second step, we assume a small perturbation $\Tilde{\F} (\E,\psi,t) = \Tilde{\F} (\E,\psi,\Omega) e^{i\Omega t}$ to the stationary function $\F(\mathcal{E})$. For $M$ equidistant bunches, the perturbation $\Tilde{\F}^l$ is characterized by a coupled-bunch mode number $l$ ($l = 0, 1, ..., M-1$), defining the phase advance $e^{-i 2\pi l/M}$between consecutive bunches, and it should satisfy the linearized Vlasov equation (see, e.g.,~\cite{Chao})
\begin{equation}
    \label{eq:vlasov_equation_new}
    \left[\frac{\partial }{\partial t} + \omega_s \frac{\partial }{\partial \psi}\right] \Tilde{\F}^l = \omega_s \frac{\partial \Tilde{U}^l_\mathrm{ind}}{\partial \psi} \frac{d \F}{d\E}.
\end{equation}
Here the perturbed induced potential $\Tilde{U}^l_\mathrm{ind}(\phi,t)$ is defined similarly to Eq.~(\ref{eq:potential}). The solution of Eq.~(\ref{eq:vlasov_equation_new}) must be periodic function of the phase $\psi$, and can be expanded in the harmonics $e^{- i m \psi}$ ($m\neq 0$). Then the corresponding harmonics of the line-density perturbation $\Tilde{\lambda}^{l}_k(\Omega)$, defined similarly to Eq.~(\ref{eq:line_density_harm}), are connected by the Lebedev equation~\cite{Lebedev1968}:
\begin{equation}
        \label{eq:matrix_lebedev}
    \Tilde{\lambda}^{l}_{k^\prime}(\Omega) = \frac{\zeta M}{h\cos\phi_{s0}} \sum_{k = -\infty}^{\infty} G_{k^\prime k} (\Omega) \, \frac{Z_k(\Omega)}{k} \, \Tilde{\lambda}^{l}_k(\Omega),
\end{equation}
with $Z_k(\Omega)=Z(k \omega_0 + \Omega)$ and the intensity parameter 
\begin{equation}
    \label{eq:xi}
     \zeta = q N_p \, h^2 \, \omega_0 \,/V_0.
\end{equation}
Note that $\Tilde{\lambda}^l_k$ are non-zero only for $k= pM + l$. The detailed derivations in variables $(\E,\psi)$ can be found, e.g., in~\cite{IK2021}.
Below we consider a stationary case with $\phi_{s0}=\pi$.  The matrix elements $G_{k^\prime k}(\Omega)$ in Eq.~(\ref{eq:matrix_lebedev}) are 
\begin{equation}
    G_{k^\prime k} = - i \omega_{s0}^2  \sum_{m=-\infty}^{\infty} m \int_0^{\E_{\max}}  \frac{d\F(\E)}{d\E} \frac{I^*_{m k}(\E) I_{m k^\prime}(\E)}{\Omega - m \omega_s(\E)} d\E,
\label{eq:BTM}
\end{equation}
and they contain functions 
\begin{equation}
    \label{eq:Imk}
    I_{m k} (\E) = \frac{1}{\pi} \int_{0}^{\pi}  e^{i\frac{k}{h}\phi(\E,\psi)} \cos m \psi \,d\psi.
\end{equation}
The coherent frequency $\Omega$ is the solution of Eq.~(\ref{eq:matrix_lebedev}) when its determinant is zero. The beam is unstable if $\text{Im}\Omega < 0$. The exact solution in a general case has to be obtained numerically, e.g., with code MELODY. Approximate analytical solutions can be also found under certain conditions.
Using a general matrix property 
\[\text{det}\left[\exp\left(\varepsilon\; X\right)\right] = \exp\left[\varepsilon\;\text{tr}\left(X\right)\right],
\]
where $\text{tr}(X)$ is the trace of a square matrix $X$, and taking into account that $X(\varepsilon) = X(0) + \varepsilon (d X/d \varepsilon)(0) + ...$,  the expansion up to the first order of $\varepsilon$ yields:
\begin{align}
    \label{eq:matrix_det}
    \text{det} & \left[I+\varepsilon\; X(\varepsilon)\right] = \text{det}\left(\exp\left\{\ln\left[I + \varepsilon X(\varepsilon) \right] \right\} \right) \nonumber \\ 
    &= \exp\left(\text{tr}\left\{\ln\left[I + \varepsilon X(\varepsilon) \right] \right\} \right)  
     \approx  1+\varepsilon\text{tr}\left[X(0)\right]. 
\end{align}
Then equating the determinant of Eq.~(\ref{eq:matrix_lebedev}) to zero, we obtain, similarly to~\cite{IK2021}, the generalized instability threshold 
\begin{equation}
    \label{eq:threshold_general}
    \zeta_\mathrm{th}(\Omega) \approx -\frac{h}{M}\left[\sum_{k=-\infty}^{\infty} G_{k k} (\Omega)\frac{Z_k(\Omega)}{k}\right]^{-1}.
\end{equation}
For the solution $\Omega=\Omega_g$, valid for any impedance model, the imaginary part of the sum is zero since $\zeta$ is a real quantity. Here we will consider the sum of NB and BB resonator impedances, $Z = Z_\NB + Z_\BB$, described by Eq.~(\ref{eq:bbr_nbr_impedance}). 
The small parameter $\varepsilon \propto \zeta_\mathrm{th}(\Omega_g)$, while its dependence on other parameters can be defined from comparison with the exact numerical solution~\cite{IK2021}.

Let us start with a more simple case of the narrow-band impedance.
If $Z_\BB=0$ and
\begin{equation}
    \label{eq:assumption2}
    \omega_{s0} \ll \Delta \omega_{\NB}  \ll M \omega_0, \,\: 
    \Delta \omega_{\NB}\ll \left|\omega_{r,\NB} - \frac{p M \omega_0}{2} \right|,
\end{equation}
where $\Delta \omega_{\NB} = \omega_{r,\NB}/2Q_\NB$ is the resonator bandwidth, all elements in Eq.~(\ref{eq:matrix_lebedev}), except with $k_{\NB}=\lfloor \omega_{r,\NB}/\omega_0 \rfloor$, can be neglected ($\lfloor x \rfloor $ denotes the rounding of $x$ to the nearest integer). Then the instability threshold is given by Eq.~(\ref{eq:threshold_general}) with only one element in the sum. Under these assumptions, the solution is exact and can be found by analyzing the equation 
\begin{equation}
    \label{eq:threshold_narrowband}
    \zeta^\NB_\mathrm{th}(\Omega) = -\frac{h}{M} \left[ G_{k_\NB k_\NB} (\Omega) \, \frac{Z_{k_{\NB}}(\Omega)}{k_{\NB}}\right]^{-1}
\end{equation}
for $\mathrm{Im}\Omega \to -0$, $\mathrm{Re}\Omega=\Omega_\NB=m\omega_s(\Tilde{\E}_m)$, $0<\Tilde{\E}_m<\E_{\max}$. The approximate analytic expression can be also obtained for a certain particle distribution from the stability diagrams~\cite{Balbekov-Ivanov1986}. 
In a short-bunch approximation, when $\phi_{\max}\ll \pi$,  we have $\E_{\max} \approx \phi^2_{\max}/2$, $\phi(\E,\psi)\approx \sqrt{2\E} \cos\psi$, $\omega_s(\E) \approx \omega_{s0}(1 - \E/8)$, and the functions $I_{mk}(\E)$ can be replaced by Bessel functions $J_m(x)$ of the first kind and the order $m$
\begin{equation}
\label{eq:Imk_approx}
I_{m k} (\E) \approx i^m J_m\left(\frac{k}{h}\sqrt{2\E} \right).
\end{equation}
The instability threshold for $\mu>1$ is the lowest for the dipole mode $m=1$~\cite{Karpov:IPAC2019}
\begin{equation}
    \label{eq:approximate_nb_threshold}
    \zeta^\NB_\mathrm{th}\approx\frac{h \phi^4_{\max} k_\NB }{16M R_\NB} \min\limits_{y \in [0,1]} \left[\frac{\left(1-y^2\right)^{1-\mu}}{ \mu (\mu+1)} J_1^{-2}\left( \frac{yk_\NB \phi_{\max}}{h}\right)\right],
\end{equation}
and it is mainly defined by the value of $R_\NB/k_\NB$, since $\mathrm{Im}Z_\NB=0$ at $\omega_r$. 

For numerical calculations, we used the parameters of the Large Hadron Collider (LHC) at injection energy from Table~\ref{tab:LHC_parameters} and binomial distribution function with $\mu = 2$ and zero-intensity $\phi_{\max}=1.3$. Due to the very large number of LHC bunches (3564), a direct comparison of macroparticle simulations with the model predictions is computationally too expensive. So we simulated nine equidistant bunches by reducing the harmonic number to nine and scaling other parameters to keep $\zeta$ and $\omega_{s0}/\omega_0$ unchanged. 

\begingroup
\begin{table}[b]
	\caption{The accelerator and rf parameters of the LHC and Super Proton Synchrotron (SPS) at $E_0=450$~GeV~\cite{HLLHCTDR, LIU2014}. 
	}
	\begin{center}
		\begin{tabular}{l  c  c  c  c }
			Parameter & Units & LHC & SPS\\
			\hline
			Circumference, $C$ & m  & 26658.86 & 6911.55\\
			Harmonic number, $h$&   & 35640 & 4620\\
			Transition gamma, $\gamma_\mathrm{tr}$ &  & 55.76 & 17.95\\
			rf frequency, $f_\mathrm{rf}$& MHz & 400.79 & 200.39\\
			rf voltage, $V_0$ & MV  & 6.0 & 7.0\\
		\end{tabular}
	\end{center}
	
	\label{tab:LHC_parameters}
\end{table}
\endgroup

For nine bunches and a single NB resonator with $k_\NB = 11$,  a  coupled-bunch mode $l=2$ with $m=1$ should become unstable above a certain threshold. Figure~\ref{fig: modes} (center) shows the dipole bunch oscillation mode ($m=1$) as a function of intensity obtained using the Oide-Yokoya method~\cite{Oide_Yokoya1990}. A mode inside the synchrotron frequency spread becomes unstable for $\zeta > \zeta^\NB_\mathrm{th}$, and the instability growth rate increases with intensity. The instability threshold found as the exact solution of Eq.~(\ref{eq:threshold_narrowband}) coincides with the emergence of the unstable mode. The solution (\ref{eq:approximate_nb_threshold}) obtained in the short-bunch approximation gives about a 10\% higher threshold, which is not surprising since the bunch with $\phi_{\max} =1.3$ is not so short.

Assuming only BB impedance, the LLD threshold can be also derived for the same distribution functions in the short-bunch approximation. Evaluating the matrix elements $G_{kk}$ for $\Omega=\Omega_\BB= \omega_s(0)$ and performing summation in Eq.~(\ref{eq:threshold_general}), one obtains~\cite{IK2021}
\begin{align}
    \zeta^\BB_{\text{th}}
    \approx 
    \frac{\pi\phi_{\text{max}}^5 }{32 \mu\left(\mu+1\right) \chi(k_\text{eff}\phi_{\text{max}}/h,\mu)}\frac{1}{\left(\text{Im}Z/k\right)_\text{eff}}.
    \label{eq:lld_threshold}
\end{align}
The function
\begin{equation} \label{eq:chi_function}
    \chi(y,\mu) =y \left[1 - {}_2 F_3 \left(1/2,1/2;3/2,2, \mu; -{}y^2 \right) \right] \nonumber
\end{equation}
is approaching $y$ for $y \gg 1$, where ${}_2F_3$ is generalized hypergeometric function.
The effective impedance is defined as 
\begin{equation}
    \label{eq:effective_impedance}
    \left(\text{Im}Z/k\right)_\mathrm{eff} = \sum_{k = -k_\mathrm{eff}}^{k_\mathrm{eff}} G_{k k} \;\mathrm{Im}Z_k/k \bigg/ \sum_{k = - k_\mathrm{eff}}^{k_\mathrm{eff}} G_{k k}
\end{equation} 
with the effective cutoff-frequency number $k_\text{eff}$ which maximizes the cumulative sum in the nominator. For a broadband resonator $k_\mathrm{eff} = k_\BB = \omega_{r,\BB}/\omega_0$.

The results of MELODY calculations for a pure BB resonator impedance with $k_\BB/h = 5$ are shown in Fig.~\ref{fig: modes} (left). The LLD mode emerges at $\zeta =\zeta^\BB_\mathrm{th}$ above the maximum incoherent frequency $\omega_s(0)$  leading to undamped but stable oscillations. Similarly to the previous case, the LLD threshold computed from the Lebedev equation agrees with the Oide-Yokoya method. The analytical approximation overestimates it by $\sim 30\%$.

\begin{figure}[tb!]
    \centering
    \includegraphics{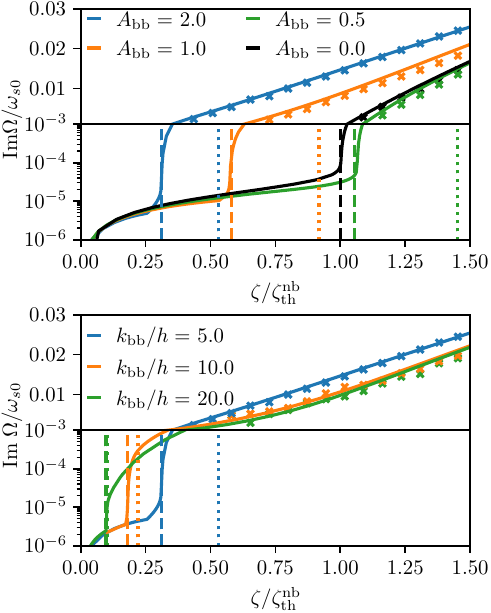}
    \caption{The growth rate of the most unstable mode versus the normalized intensity parameter $\zeta$  for various strengths (top) and cutoff frequencies (bottom) of the BB impedance. Instability and LLD thresholds found from exact Eq.~(\ref{eq:matrix_lebedev}) are shown with vertical dashed and dotted lines, respectively. BLonD simulations are marked with crosses. The parameters from Table~\ref{tab:LHC_parameters} scaled to $h=9$; $k_\NB/h = 11/9$, $Q_\NB = 100$, $R_\NB = 37$~kOhm, $Q_\BB = 1$, 
    and $R_\BB =3 A_\BB \times (k_\BB/h)$ kOhm. Top: $k_\BB/h = 5$, bottom: $A_\BB = 2$.}
    \label{fig:role_bbr}
\end{figure}

Finally, if we combine the BB and NB impedance contributions, the  threshold~(\ref{eq:threshold_general}) can be written in the form
\begin{equation}
    1 / \zeta_{\mathrm{th}}(\Omega_g) = 1/\zeta^\NB_{\mathrm{th}}(\Omega_g) + 1/\zeta^\BB_{\mathrm{th}}(\Omega_g).
    \label{eq:general}
\end{equation}
The coherent mode $\Omega_g$ differs from $\Omega_\NB$ and $\Omega_\BB$ found for each impedance separately. Nevertheless, for a first estimate, one can use $\zeta^\NB_{\mathrm{th}}(\Omega_\NB)$ and $\zeta^\BB_{\mathrm{th}}(\Omega_\BB)$ in Eq.~(\ref{eq:general}). If $\zeta^\BB_{\mathrm{th}}(\Omega_\NB) \ll \zeta^\NB_{\mathrm{th}}(\Omega_\BB)$, $\Omega_g \approx \Omega_\BB$ and the actual CBI threshold is reduced, as shown in Fig.~\ref{fig: modes} (right), where it is even below the LLD threshold. In the opposite case, the BB impedance has a negligible impact and $\Omega_g \approx \Omega_\NB$. 
The relative role of each contribution in Eq.~(\ref{eq:general}) can also be seen in Fig.~\ref{fig:role_bbr} from instability thresholds numerically obtained for different $(\text{Im}Z/k)_\mathrm{eff}$ and $k_\mathrm{bb}$ using Eq.~(\ref{eq:matrix_lebedev}). As expected, a larger BB impedance leads to a lower CBI threshold, except for the case when $\zeta^\BB_{\mathrm{th}}(\Omega_\BB) \gg \zeta^\NB_{\mathrm{th}}(\Omega_\NB)$. BLonD simulations follow closely the growth rates obtained using the Oide-Yokoya method and the direct solution of the Lebedev equation confirms the thresholds. 

\begin{figure}[bt]
    \centering
    \includegraphics{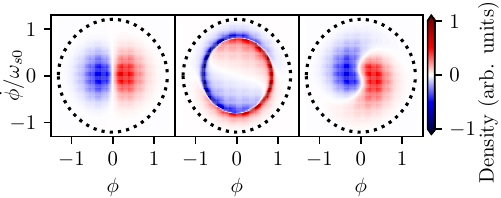}
	\caption{Perturbed particle distribution function $\Tilde{\F}(\phi,\dot{\phi})$ in phase space for LLD mode (BB impedance) at $\zeta/\zeta^{\NB}_\mathrm{th} = 0.5$ (left), unstable mode driven by NB impedance at $\zeta/\zeta^{\NB}_\mathrm{th} = 1.1$ (center), and for unstable mode with NB and BB impedances at $\zeta/\zeta^{\NB}_\mathrm{th} = 0.5$ (right). The dotted line is the outermost particle trajectory in the bunch. Other parameters are as in Fig.~\ref{fig: modes}.
 }
	\label{fig:modes_phase_space}
\end{figure}

The impact of $k_\mathrm{eff}$ is shown in Fig~\ref{fig:role_bbr} (bottom). The CBI threshold is reduced for larger $k_\BB$ since the LLD threshold is also reduced, and its effect on the overall instability threshold is increased. 
We see, however, that the growth rates above a certain intensity become smaller for larger $k_\BB$. 
For this new instability mechanism, the mode is localized in the bunch center (Fig~\ref{fig:modes_phase_space}, right), similar to a single-bunch LLD (Fig~\ref{fig:modes_phase_space}, left). For larger $k_\BB$, the mode spectrum shifts towards higher frequencies as for LLD mode~\cite{IK2021} and interacts weaker with NB impedance since $k_\NB<k_\BB$. 
For $Z_\BB =0$, the perturbation looks very different and involves mainly high-amplitude particles for the considered $k_\NB$ (Fig~\ref{fig:modes_phase_space}, center).

Considering LHC parameters without scaling (Table~\ref{tab:LHC_parameters}), in the full ring, the lowest CBI threshold due to a higher-order mode (HOM) of the future crab cavities~\cite{HLLHCTDR} with $R_{\NB}=280$~kOhm, $f_{r,\NB} =582$~MHz, and $Q_{\NB} = 1360$ is three times smaller with BB impedance~\cite{karpov:ipac2022}. 

\begin{figure}[tb!]
    \centering
    \includegraphics{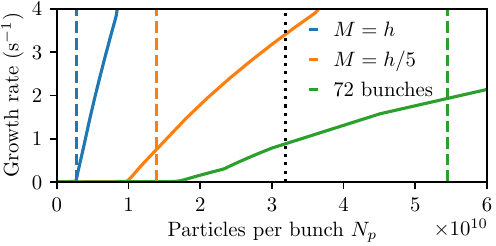}
    \caption{ Growth rates of the most unstable mode (solid lines) versus bunch intensity found with MELODY for SPS with $M=h$ (blue), $M=h/5$ (orange) and LHC train of 72 bunches spaced by $5/f_\mathrm{rf}$ (green) with corresponding CBI thresholds driven by only HOM at 914 MHz (dashed lines). The LLD threshold is shown by a black dotted line. The SPS parameters from Table~\ref{tab:LHC_parameters}, zero-intensity $\phi_{\max}=0.9$, and $\mu=1.5$.}
    \label{fig: SPS}
\end{figure}

The SPS, being the LHC injector, also provides beams for the fixed-target physics. For these beams filling the whole ring, CBI is driven by a higher-order mode (HOM) in the main 200 MHz rf system with $f_r \approx 914$ MHz~\cite{Shaposhnikova1999}. The CBI thresholds found for the full SPS impedance model~\cite{SPS_impedance, LasheenPRAB:2017} and for only HOM impedance practically coincide (see Fig.~\ref{fig: SPS}) since the LLD threshold is higher by an order of magnitude. If the beam fills every 5th rf bucket (LHC-type), the HOM-driven CBI threshold is higher and the resulting threshold is more affected by the BB part of the SPS impedance. This explains why the LHC-type bunch train in the SPS has a very low CBI threshold, which also weakly depends on the number of bunches in a train~\cite{karpov:ipac2022}. To deliver high-intensity LHC beams, an additional 800-MHz rf system leading to the raised CBI threshold is routinely deployed~\cite{LIU2014}. 

To summarize, we proposed the approach to analyze beam stability in the presence of both broadband and narrowband impedance sources. The broadband impedance can significantly reduce the threshold of coupled-bunch instability driven by the narrowband impedance, and there is a new instability mechanism associated with it. The derived generalized analytical expression shows the key role of LLD and demonstrates how two impedance contributions add up. For the LLD-dominated case, the values of the effective BB impedance and its cutoff frequency are important. The main conclusions are verified by macroparticle simulations and they are consistent with beam observations in the SPS. This understanding can help in finding mitigation measures aimed at increasing Landau damping for existing high-current synchrotrons. The discovered effect should be also taken into account in the design of future rings.

\begin{acknowledgments}
We are grateful to Heiko Damerau for his useful comments.
\end{acknowledgments}

\nocite{apsrev41Control}
\bibliographystyle{apsrev4-1}
\bibliography{main}

\end{document}